\documentclass[aps,prl,preprint,groupedaddress]{revtex4}

\usepackage{graphicx}

\begin{document}


\title{High Field de Haas - van Alphen Studies of the Fermi Surfaces 
of LaMIn$_{5}$ (M = Co, Rh, Ir)}


\author{Donavan Hall}
\author{Luis Balicas}
\affiliation{National High Magnetic Field Laboratory, Florida State University, Tallahassee, FL 32306}
\author{Z.\ Fisk}
\affiliation{Department of Physics and Astronomy, University of California, Irvine, CA 92697}
\author{R.\ G.\ Goodrich}
\affiliation{Department of Physics and Astronomy, Louisiana State University, Baton Rouge, LA 70803}
\author{U. Alver}
\affiliation{Department of Physics, Kahramanmaras Sutcu Imam University, 46100, K.Maras, Turkey }
\author{J.\ L.\ Sarrao}
\affiliation{Los Alamos National Laboratory, Los Alamos, NM 87545}


\date{\today}

\begin{abstract}
    We report measurements of the de Haas - van Alphen effect on a
    series of compounds, LaMIn$_{5}$ (M = Co, Rh, Ir).  The results show
    that each of the Co and Ir Fermi surfaces (FSs) exhibit some
    portions that are two dimensional and some portions that are three
    dimensional.  The most two dimensional character is exhibited in
    LaCoIn$_{5}$, less two dimensional behavior is seen in LaIrIn$_{5}$, no part
    of Fermi surface of LaRhIn$_{5}$ is found to have a two dimensional
    character.  Thus the two dimensionality of portions of the FSs is
    largely determined by the d character of the energy bands while
    all of the effective masses remain $\leq$ 1.2.  This fact has
    implications for the causes of the heavy fermion nature of
    superconductivity and magnetism in the Ce-based compounds having
    the similar  composition and structure.  All of the
    measurements were performed at the National High Magnetic Field
    Laboratory using either cantilever magnetometry or field
    modulation methods.
\end{abstract}


\maketitle

\section{Introduction}

Prior to the discovery of magnetic superconductors it was believed
that the inclusion of magnetic rare-earth atoms in a material would
effectively prevent the formation of a superconducting state.  Weak
ferromagnetic ordering was enough to stop the formation of Cooper
pairs.

In the mid-seventies rare-earth ternary compounds were
discovered that had both antiferromagnetic ordering and a
superconducting state.\cite{Fisher1975,Fertig1977}  In these materials
ferromagnetic ordering via exchange interactions broke apart Cooper
pairs; however, long range antiferromagnetic order was essentially invisible to
the formation of superconducting charge carriers.

Not long after the discovery of these antiferromagnetic superconductors, the
first heavy fermion superconductors were discovered.\cite{Steglich1979}
These materials also exhibit weak antiferromagnetic ordering
with the addition that the effective mass of the charge carriers are
much larger than in normal metals, typically 10 to a 100 times the
electron mass.

In order to fully explore the probable causes of these measured
phenomena, it is necessary to find clean, isostructural systems where
the effects of chemical doping changes can be measured along with
other thermodynamic changes.  One such system is the Rare Earth-115 family
consisting of dozens of compounds and intermediate dopings.

For the present work we have focused on one subset of this RE-115
family, the REMIn$_{5}$s (RE = La or Ce, M = Co, Rh, or Ir).  One of the
most interesting members of this family is CeCoIn$_{5}$, since the
transition from the superconducting state to normal state at high
magnetic fields is first order which offers the possibility that this
material is the first to exhibit an FFLO superconducting 
state.\cite{Bianchi2003}

CeRhIn$_{5}$ has an antiferromagnetic ground state that can be driven into
a superconducting state with the application of pressure.\cite{Shishido2006}
In CeIrIn$_{5}$ the resistive and magnetic transition in zero field do not 
occur at the same temperature.

Gauging to what extent the 4f electrons contributed by the Ce atoms
determine the various properties of these materials requires the slow
removal of these electrons from the lattice.  Measurements on
Ce$_{x}$La$_{1-x}$-115s were the subject of another study.  It was found that
even 10 percent dilutions of the magnetic 4f electrons would
effectively remove the heavy fermion and superconducting properties of
these materials.  The conclusion is that magnetism is at
the heart of understanding the rich physics of this system.

To date there has only been a single published study of the Fermi
surface of the LaMIn$_{5}$ materials and that study only concerned LaRhIn$_{5}$.
It was found (and we confirm this) that the CeRhIn$_{5}$ FS is almost
identical to LaRhIn$_{5}$.  This is taken as an indication that the
f-electrons are localized in CeRhIn$_{5}$ and thus are not a major
contributor to the FS topology.

CeCoIn$_{5}$ and CeIrIn$_{5}$ on the other hand are thought to exhibit some
delocalization of the f-electrons.  This would suggest that the
removal of the f-electrons from the material would have a significant
affect on the FS. We have measured the dHvA effect in LaCoIn$_{5}$ and
LaIrIn$_{5}$.  We find that while there are some differences, the general
character of the FS is preserved under the complete removal of the
f-electron.  This indicates that it is the d-electrons in these
materials that contribute most to the FS with the f-electrons
contributing only small changes and a significant renormalization of
the effective mass of the conduction electrons.

Because electrons on the Fermi surface (FS) are thought to play a role
in all of these exotic properties, there has been considerable effort
made in measuring their FSs over the past few years.  The first
measurements on the CeMIn$_{5}$ series were done on the Co based 
material\cite{Hall2001a},
next the Rh material\cite{Hall2001b}, and then the Ir 
material.\cite{Haga2001}  In every case there
is evidence that at least parts of the FSs are cylindrical with axes
along the $\Gamma$Z direction of the Brillouin Zone, similar to
high-temperature superconducting cuprates.  Both CeMIn$_{5}$ and LaMIn$_{5}$ can
be described as having a layered structure with the layers viewed as
alternating La-In planes and M-In planes.  If electrical conduction
only takes place in two dimensions in a material the electronic
structure of the material is confined to these two dimensions and the
FS will be perfectly cylindrical.  Consequently, the dHvA measurements
that determine the cross sectional areas of the FS perpendicular to an
applied field would have an angular dependence of the measured
frequencies proportional to 1/cos($\theta$)  where $\theta$  is the angle between
the direction of the applied field and the direction perpendicular to
the layers.  When there is conduction between the planes other 3D
pieces of FS can exist, and the nearly cylindrical FS will undulate in
area along the $\theta$  = 0 direction, and several frequencies may appear to
obey the 1/cos($\theta$)  rule over a limited angular range.

It is known that the heavy fermion properties of Ce based compounds
are due to delocalization of the Cerium 4f electrons forming bands
with f character.  However, the delocalization may not be complete,
and energy band calculations of the real ground state only recently
have been performed.  For this reason, we have undertaken the
measurements of the FS of the La based LaMIn$_{5}$ (M = Co, Rh, Ir)
compounds that have the same structure as the Ce based ones, but have
no 4f electrons.  The results give information about the effects of
the different d bands and a starting point for future energy band
calculations that should be able to describe these FSs correctly.

\section{Experimental Procedure and Data Analysis}

All of the results reported here are from data taken at the National
High Magnetic Field Laboratory (NHMFL) in Tallahassee, FL. Most of the
measurements were made using a metal film cantilever in a rotating
sample holder.  However, some of the results were checked using
balanced pickup coils and magnetic field modulation.  (These 
techniques are described in many previous papers, including our own, 
e.g. see \cite{Hall2001a}.)  The frequency of
the dHvA effect was measured for two different samples of each of the
three values of M (Co, Rh, Ir) using different measurement techniques.
Field modulation using balanced pickup coils is more sensitive to the
high frequencies, while the metal film torque measurement is best for
detecting the lowest frequencies.  The samples, and the cantilever or
pick up coils were immersed either in liquid $^{4}$He that could be pumped
to 1.5 K or in liquid $^{3}$He with a base temperature of approximately 0.5
K. All of the data analyzed was performed on data taken in the field
range of 15 to 33 T in one of the resistive magnets at the NHMFL.

\begin{figure}
\includegraphics{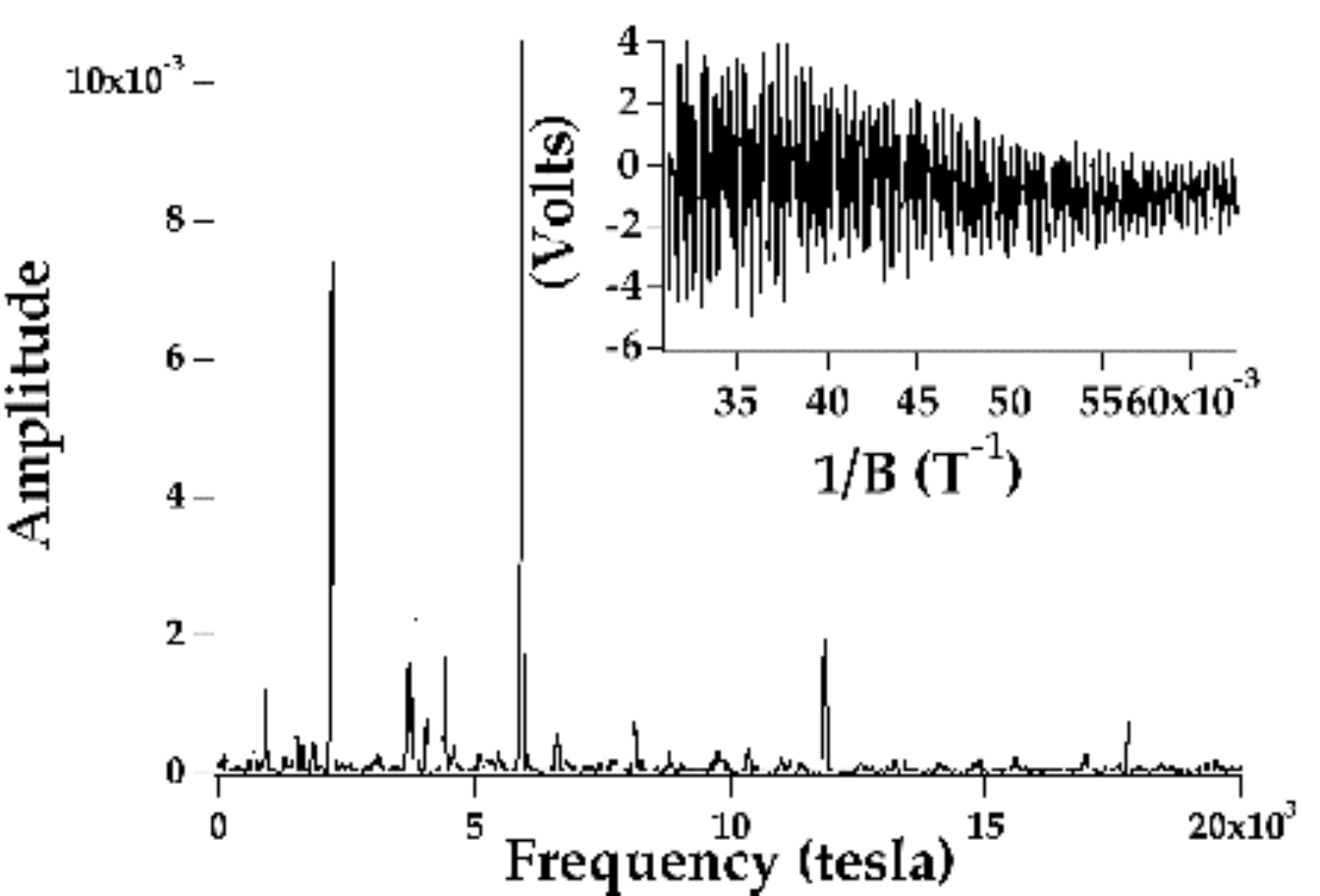}%
\caption{\label{LaIrIn5sigs} Fourier transform of the dHvA data for
LaIrIn$_{5}$.  The raw data plotted as a function of reciprocal field is
shown in the inset.}
\end{figure}

In Figure \ref{LaIrIn5sigs} we show data from LaCoIn$_{5}$ with the field perpendicular to
the planes along the c axis.

In the main graph a Fourier analysis of the data for fields between 15
and 33 T is shown, while in the inset the actual raw data over this
field range is shown.  The quality of this data is typical for all of
the samples investigated.  As can be seen there are multiple
frequencies, some of which are harmonics of fundamental frequencies,
and some are fundamental frequencies plus or minus another fundamental
frequency due to magnetic interactions, i.e. the B field seen by
electrons on one extremal area orbit is modulated due to oscillations
of electrons on other extremal area orbits.  In Fig.1 the peaks in the
Fourier transform at 12 an 18 kilo-Tesla (kT) are the second and third
harmonics of the peak at 6 kT. Each reported fundamental frequency
given below has been checked to assure it is not either of the
non-fundamental frequencies described above.

\section{Results and Discussion}

\begin{figure}
\includegraphics{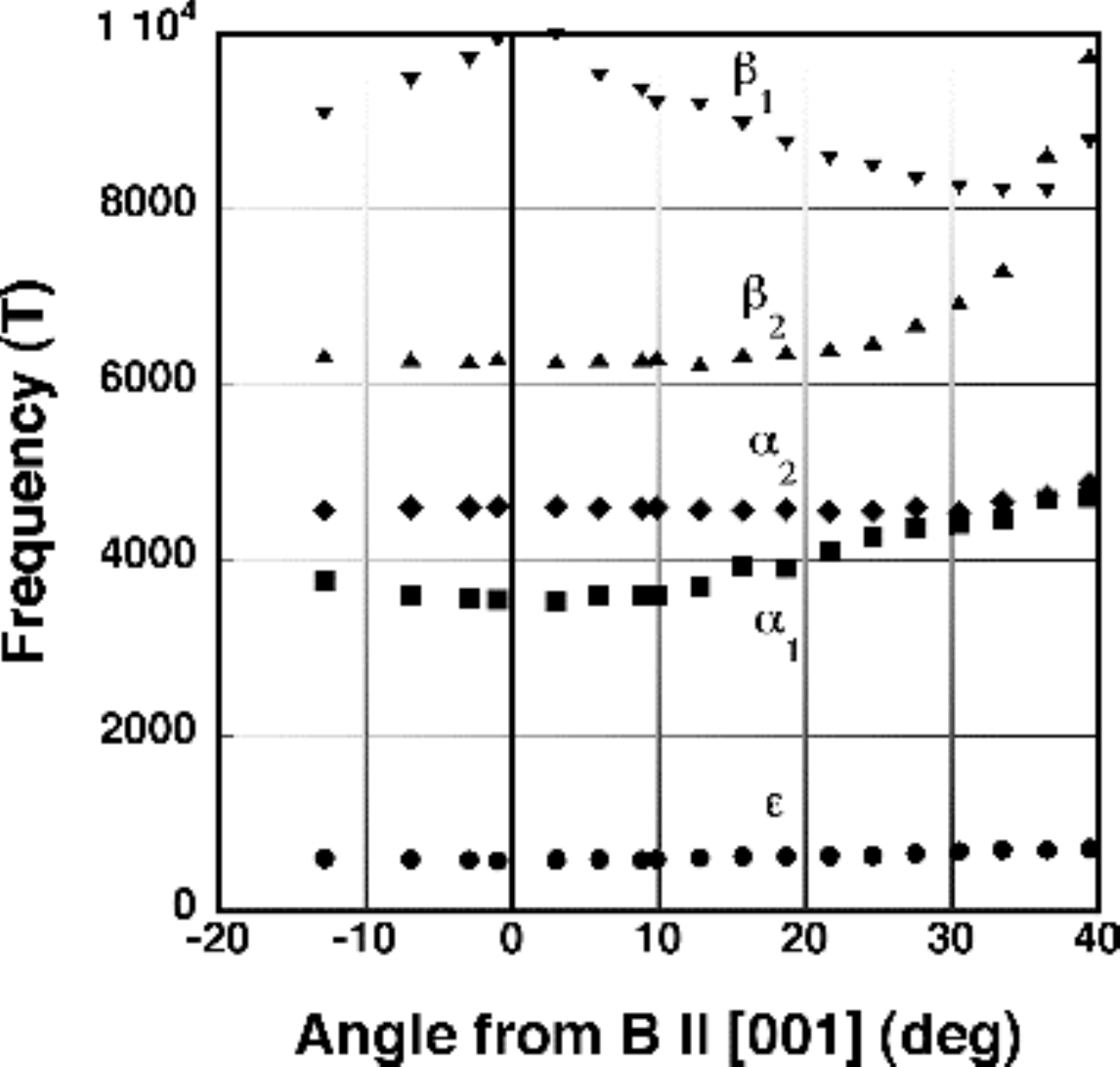}%
\caption{\label{LaRhIn5rot} dHvA frequencies vs.  angle between [001]
and the field direction for LaRhIn$_{5}$.}
\end{figure}

\begin{figure}
\includegraphics{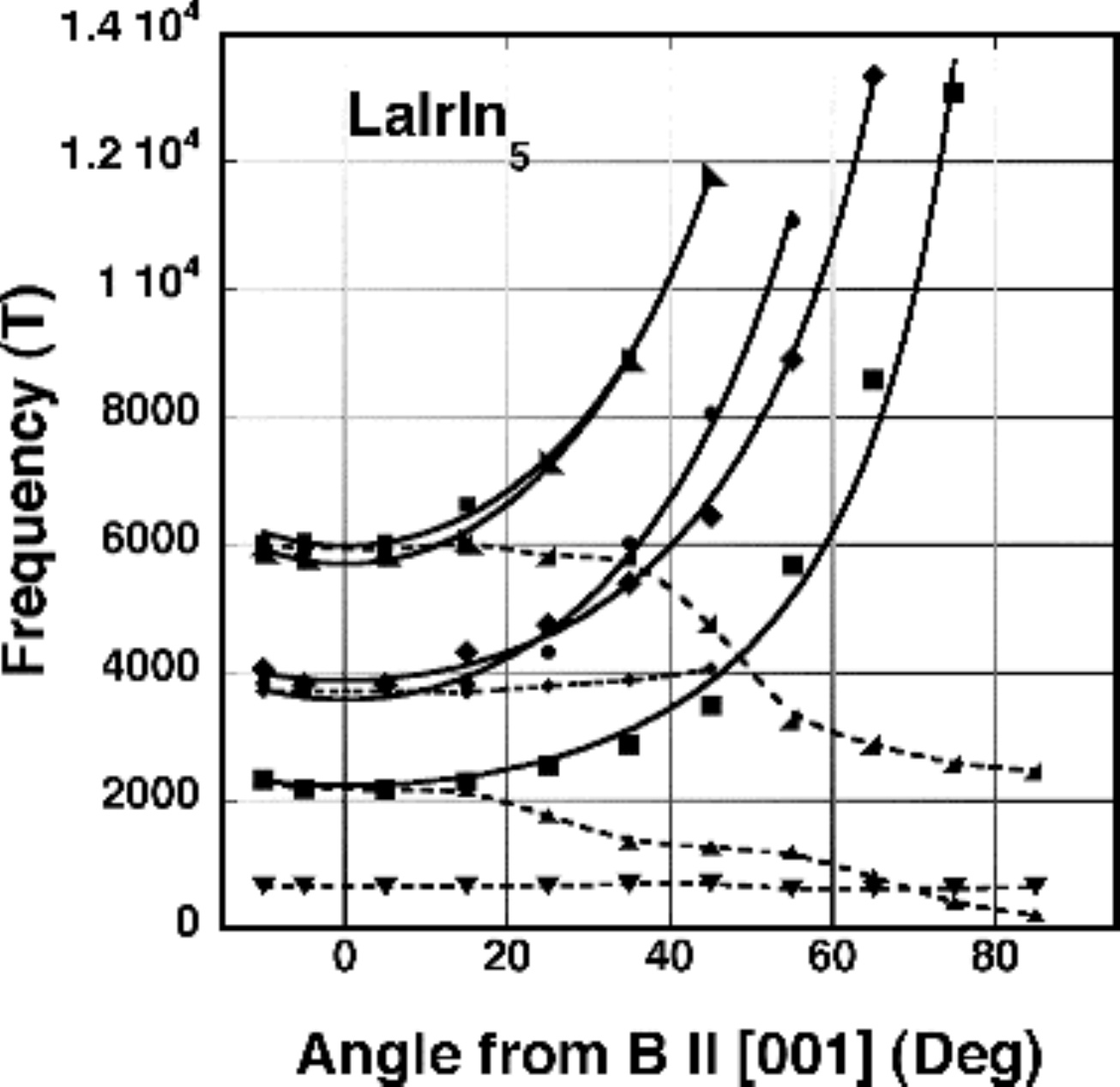}%
\caption{\label{LaIrIn5rot} dHvA frequencies vs.  angle between [001]
and the field direction for LaIrIn$_{5}$.}
\end{figure}

\begin{figure}
\includegraphics{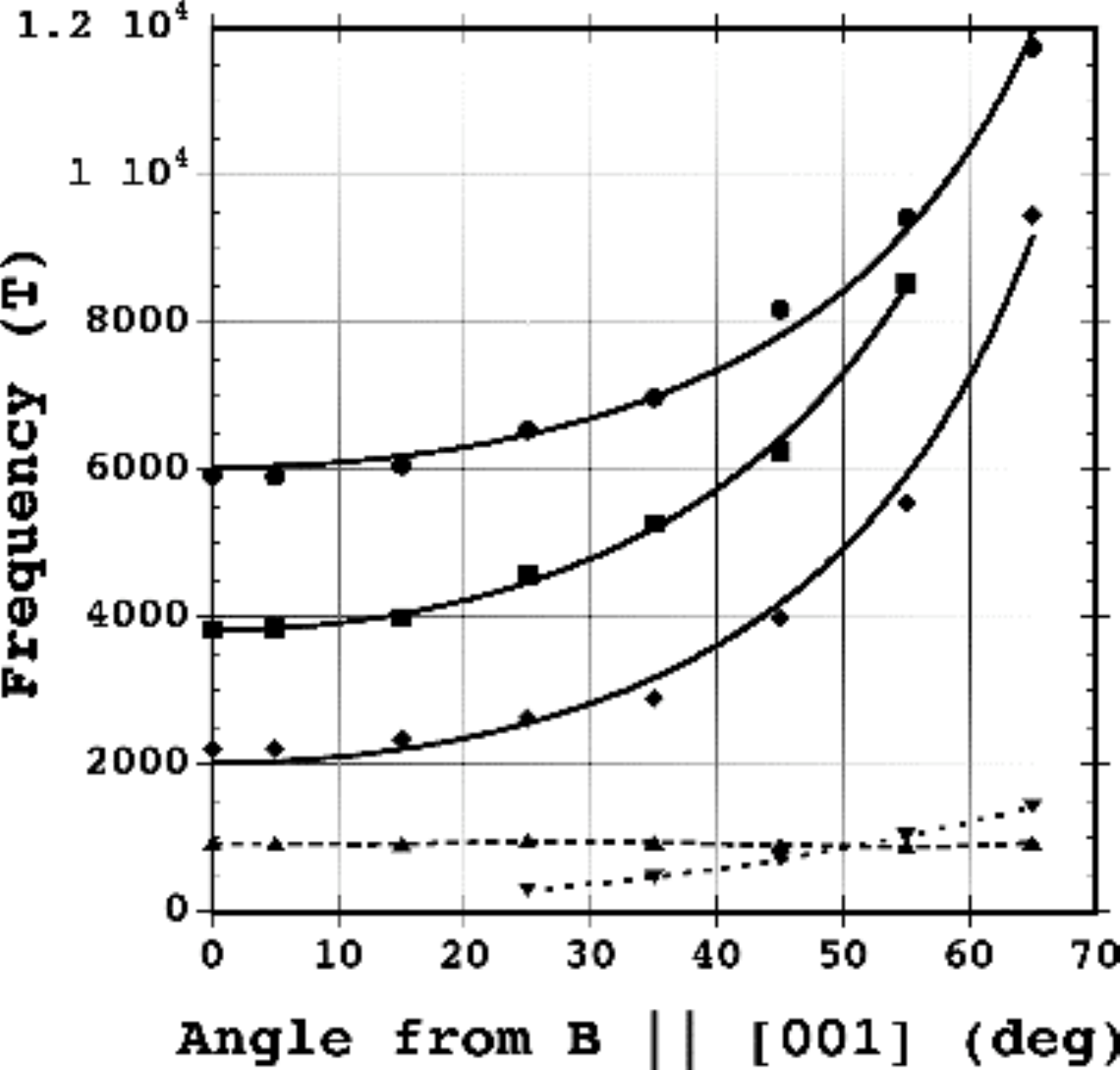}%
\caption{\label{LaCoIn5rot} dHvA frequencies vs.  angle between [001]
and the field direction for LaCoIn$_{5}$.}
\end{figure}

The results of the angular dependent measurements on the three LaMIn$_{5}$
compounds are shown in Figures \ref{LaRhIn5rot}, \ref{LaIrIn5rot}, 
and \ref{LaCoIn5rot} for M = Rh, Ir, and Co
respectively.

The first observation is that the number of frequencies observed
increases as one proceeds from 3\textit{d} (Co) to 4\textit{d} (Rh) then to 5\textit{d} (Ir)
contributions to the conduction bands.  This result indicates that
there are different interactions between the \textit{d} band electrons and the
free electron \textit{s} band states depending on increasing angular momentum
of the \textit{d} electrons.  The first results we obtained were for LaRhIn$_{5}$
with the measured frequencies as a function of angle shown in Figure
\ref{LaRhIn5rot}.  The consequences of the very low frequency (7 T) that we have
reported on previously\cite{Goodrich2002} is not discussed in this paper.  Overall these
results are consistent with those obtained by Shishido \textit{et 
al.}\cite{Shishido2001}  In the
Rh case band calculations can be done accurately displaying remarkable
agreement with both the experimental results reported in Ref.  \cite{Shishido2001} as
pointed out in Ref.  \cite{Shishido2006}.  In Figures 
\ref{LaIrIn5rot} and \ref{LaCoIn5rot} the solid lines are fits
to f = f$_{0}$/cos($\Theta$), where f$_{0}$ is the frequency at $\Theta$  = 0, over limited
angular ranges.  The frequencies that are quasi-cylindrical all arise
from extremal area orbits on the band 15 - electron part of the FS,
normally denoted as $\alpha_{1}$, $\alpha_{2}$, and $\alpha_{3}$.  We note that while cylindrical
like orbits are seen in LaCoIn$_{5}$ and LaIrIn$_{5}$, they are more separated
than those seen in CeCoIn$_{5}$\cite{Hall2001a} and CeIrIn$_{5}$\cite{Haga2001} indicating that the f
electrons also cause this piece of the FS to become more cylindrical.
The observation to be made here is that the overall interaction of the
\textit{d} levels of the M in the CeMIn$_{5}$ compounds is significant and the Ce f
levels are not the only electron states involved in determining the
quasi-cylindrical nature of the FS shapes.  They do, however cause
increases in effective mass in each case.  As was previously shown the
partially occupied f bands for the Co and Ir Ce based compounds cause
increases in the FS dimensions, while the Rh based Ce and La FSs
remain nearly identical\cite{Harrison2004}.

\begin{acknowledgments}
    
The work at the NHMFL was performed under the auspices of the National
Science Foundation, and the State of Florida.  R. G. G. was supported
directly by the NSF while Z. F. acknowledges Grant No.
NSF-DMR-0503361.

\end{acknowledgments}

\bibliography{re115}

\end{document}